# AN OPTICAL SCHEME FOR QUANTUM MULTI-SERVICE NETWORK

FÁBIO ALENCAR MENDONÇA[1,2], DANIEL BARBOSA de BRITO[1] and RUBENS VIANA Ramos[1]

fabioalencar@ifce.edu.br danielbb@deti.ufc.br rubens@deti.ufc.br

[1]*Lab. of Quantum Information Technology, Department of Teleinformatic Engineering – Federal University of Ceara - DETI/UFC, C.P. 6007 – Campus do Pici - 60455-970 Fortaleza-Ce, Brazil.*
[2] *Federal Institute of Education, Science and Technology of Ceara, Fortaleza-Ce, Brazil*

Several quantum protocols for data security having been proposed and, in general, they have different optical implementations. However, for the implementation of quantum protocols in optical networks, it is highly advantageous if the same optical setup can be used for running different quantum communication protocols. In this direction, here we show an optical scheme that can be used for quantum key distribution (QKD), quantum secure direct communication (QSDC) and quantum secret sharing (QSS). Additionally, it is naturally resistant to the attack based on single-photon detector blinding. At last, we show a proof-of-principle experiment in 1 km optical fiber link that shows the feasibility of the proposed scheme.

## 1. Introduction

The field of secure quantum communication has been intensively investigated in recent decades. Although there are already several quantum protocols for data security [1-6], up to this moment only quantum key distribution became commercially available. Considering the implementation of different quantum communication protocols in optical networks, a highly desirable situation would be the development of an optical setup able to run different quantum protocols, integrating, in a simple way, different quantum services in the same optical network. In this direction, this work presents an optical scheme, based on thermal and coherent states, which can support QKD, QSDC and QSS protocols. Furthermore, the proposed scheme can be easily implemented and it is resistant against intercept-resend, beam splitter and photon number splitting attacks, as well the external control of single-photon detectors [7].

Before explaining the proposed optical scheme, we give a brief review of coherent and thermal states. These states are described by the following density operators

$$\rho_\alpha = |\alpha\rangle\langle\alpha|,\quad |\alpha\rangle = \sum_{n=0}^{\infty} \exp\left(-\frac{|\alpha|^2}{2}\right)\frac{\alpha^n}{\sqrt{n!}}|n\rangle \tag{1}$$

$$\rho_t = \frac{\mu_t^n}{(1+\mu_t)^{1+n}}|n\rangle\langle n|. \tag{2}$$

In (1) and (2), $\mu_t$ and $|\alpha|^2$ are, respectively, the mean photon number of the thermal and coherent states. Their distinguishability can be measured by the overlap

$$\langle\alpha|\rho_t|\alpha\rangle = \exp\left[-|\alpha|^2/(1+\mu_t)\right]\Big/(1+\mu_t). \tag{3}$$

In particular, if $\mu_t=|\alpha|^2$ then the lower the mean photon number the worse is the distinguishability. However, if $\mu_t=|\alpha|^2\neq 0$, the states $\rho_\alpha$ and $\rho_t$ can be distinguished if one has a large enough number of samples of one of them. This task can be realized by a threshold single-photon detector. The probabilities of thermal and coherent states to fire an avalanche in a threshold single-photon detector are (not coinsidering the afterpulsing), respectively

$$P_t = 1 - \frac{1}{1+\eta\mu_t}(1-p_d) \quad (4)$$

$$P_c = 1 - \exp\left(-\eta|\alpha|^2\right)(1-p_d). \quad (5)$$

In (4) and (5), $\eta$ and $p_d$ are, respectively, the single-photon detector quantum efficiency and dark count probability.

Measuring with a spectrum analyser, in a fixed frequency band, the electrical power of the signal produced by a threshold single-photon detector, a large sample of thermal states can be fully distinguished of a large sample of coherent states when both of them have the same low mean photon number. This happens because the electrical power in a fixed band is proportional to $P–P^2$, where $P$ is the probability of an avalanche to be fired [8]. Since the probabilities in (4) and (5) are different when $\mu_t$=$|\alpha|^2$ ($P_c > P_t$), the electrical powers measured will also be different.

## 2. Optical setup for quantum multi-service network

The proposed optical scheme is shown in Fig. 1. The goal of this setup is to use the thermal states to protect the coherent states emitted by Alice and phase modulated by Bob.

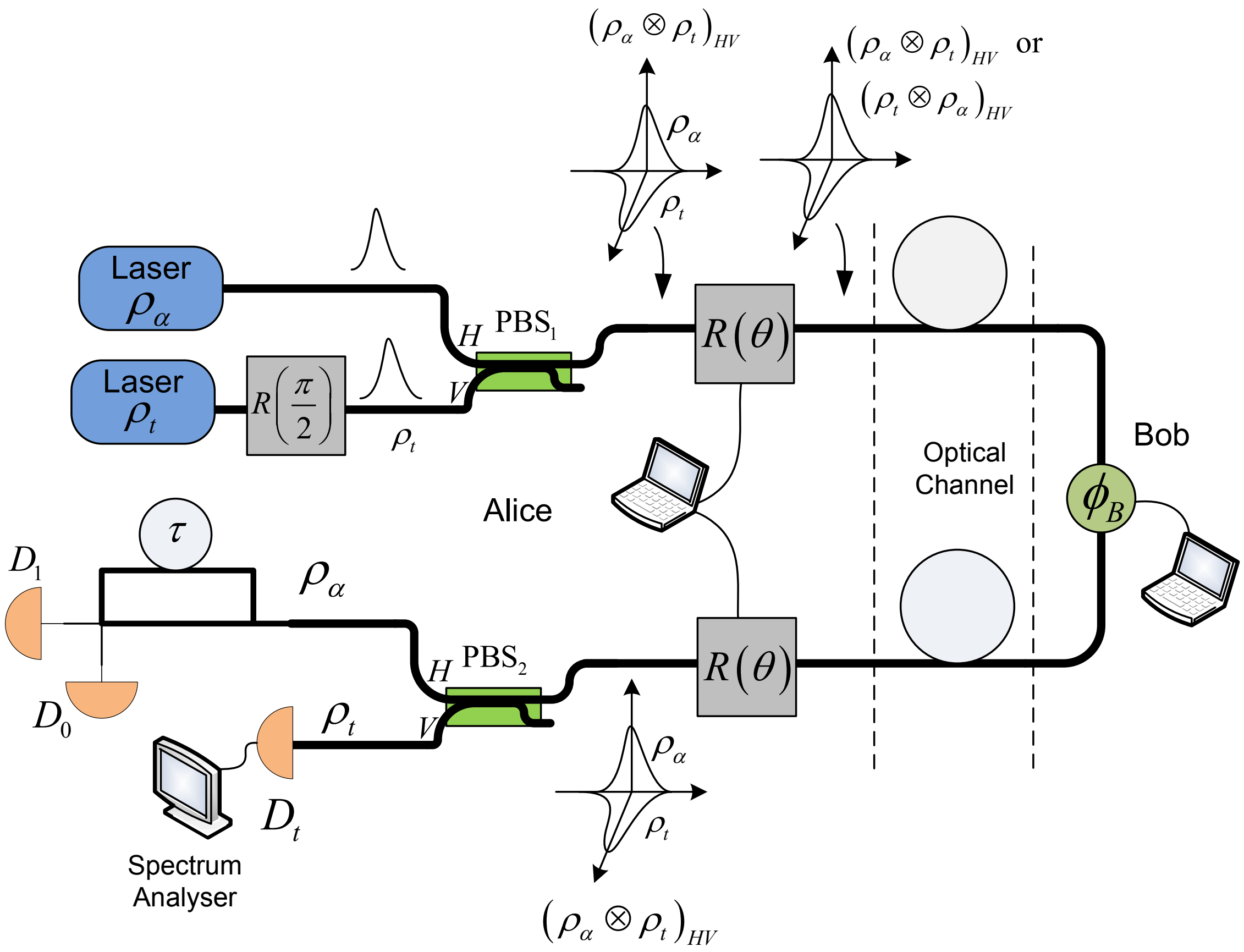

Fig. 1 - Optical setup for implementation of a multi-service quantum network. PBS – polarizing beam splitter; R – polarization rotator; $\phi_B$ – phase modulator.

Basically, Alice produces optical pulses having a coherent state at the horizontal mode and a thermal state at the vertical mode, both of them having the same low mean photon number. Following, Alice, randomly, sets her polarisation rotator $R(\theta_1)$ in 0 or $\pi/2$. Thus, for each optical pulse at Alice's output, the quantum state entering the optical channel is $1/2(\rho_\alpha \otimes \rho_t)_{HV}$ + $1/2(\rho_t \otimes \rho_\alpha)_{HV}$. Bob, by its turn, has a polarisation insensitive phase modulator [9]. He codes his information in the pulses sent by Alice by applying a phase-shift. Bob's phase modulation does not change the thermal state but it adds the phase $\phi_B$ to the coherent state. Now, the optical pulses are sent back to Alice. For each pulse arriving, she applies the same polarisation rotation she had applied when the pulse was leaving her place, that is $\theta_2=\theta_1$. Thus, before $PBS_2$, all pulses will have the coherent state at $H$-mode and thermal state at $V$-mode. These modes are separated by $PBS_2$ and the thermal state at the $V$-mode is monitored by a single-photon detector plugged to a spectrum analyser that will measure the electrical power in a fixed band. On the other hand, the coherent state at $H$-mode is sent to a fibre interferometer whose time difference between upper and lower arms, $\tau$, is equal to the time separation between two consecutive pulses.

## 3. Security analysis

The security of the proposed optical setup can be explained as follows: Alice has two secrets: the mean photon number used and the polarisation rotations applied. During the communication, only one mean photon number value is used, but only Alice knows its value. As explained before, the coherent and thermal states with the same mean photon number can be distinguished if one has a large number of samples. With the setup shown in Fig. 1, only Alice can have a large amount of samples (pulses) because she is the only one able to separate with 100% of certainty the coherent and thermal states. Thus, Alice knows exactly the electrical power value she has to measure at the thermal state output. If she measures a different value, she will know that at thermal state output there is another type of quantum state or the mean photon number of the thermal state was changed.

In order to gain some information, the eavesdropper, Eve, has to attack the coherent state after Bob's phase modulation. However, Eve does not know in which polarisation mode it is, hence, she has to attack both modes. In the intercept-resend attack, having only one pulse to make correctly the distinguishability, sometimes the eavesdropper will be confused and, during the state reconstruction and, with some probability (depending on the method used to determine in which mode is the coherent state), she may change $\rho_\alpha$ and $\rho_t$ positions. In this case, Alice will receive some coherent states at the thermal state output and some thermal states at the coherent state output. Alice will notice this attack because the electrical power value measured at thermal state output will be different from the expected value. Moreover, the thermal states at coherent state output will increase the error rate of the quantum communication protocol.

For the photon number splitting (PNS) attack, Eve will have to count the photon number of both modes. If each mode has at least two photons, a single-photon from each mode is captured and the rest of the photons are sent to Alice through a lossless fibre. If at least one of the modes has only one or zero photons, the optical pulse is absorbed and a vacuum state is sent to Alice. As can be seen, in this attack Eve's action does not causes the appearance of coherent states at the thermal output, but it changes the photon number distribution of the pulses arriving at the single-photon detector at thermal output and, hence, the electrical power value measured at thermal state output once more will be different from the expected value. In order to see this clearly, let $p^{\alpha}_{0(1)}$ and $p^{t}_{0(1)}$ be, respectively, the probabilities of the coherent and thermal states sent by Alice having zero (one) photon. Thus, the quantum state of the light arriving at Alice's place in the $V$-mode is, approximately, $(1\text{-}q)|0\rangle\langle 0|+q|1\rangle\langle 1|$, where $q = \left[1 - p^{\alpha}_{0(1)} - p^{\alpha}_{0(1)}\right]\left[1 - p^{t}_{0(1)} - p^{t}_{0(1)}\right]$. As can be noted, for simplification, the situations where the pulses sent by Alice have more than two photons were not considered since the mean photon number used is much lower than 1. Thus, the probability of detection caused by that state is $1\text{-}(1\text{-}q\eta)(1\text{-}p_d)$. In order to do not disturb the electrical power value measured by Alice, the condition $1\text{-}(1\text{-}q\eta)(1\text{-}p_d)=1\text{-}(1\text{-}p_d)/(1+\eta\mu_t)$ must be obeyed. However, for $\mu_t<10$ this condition is never satisfied for any value of $\eta$, hence the PNS attack will cause an error in Alice.

The beam splitter attack can be realized without disturbing Alice's measurement if the beam splitter used has reflectance equal to the channel losses and Eve provides a lossless channel between her place and Bob's place. However, since Eve cannot attack all pulses, the amount of information obtained by Eve is limited. Like the PNS attack, she has to obtain at least a photon from each mode.

At last, the optical setup shown in Fig. 1 is naturally resistant to the attack in which the single-photon detectors are externally controlled by Eve by using strong light [5]. If Eve tries to control Alice's single-photon detectors the strong light will change the electrical power measured by Alice at the thermal state output, indicating that an attack is happening.

For all attacks discussed up to now we were concerned only with the probability of Eve to cause an alert signal in Alice by changing the electrical power value measured. However, even when this is not the case, Eve still has a hard problem to solve: she does not know which mode ($H$ or $V$) contains the useful information, hence, she has to measure both of them and try to discover what is the useful information. For example, if the information is coded in the difference of phase between two consecutives coherent states (as it will be discussed latter), and Eve has success in her attack getting four photons from the two consecutive pulses, namely $ph_{1c}$, $ph_{1t}$, $ph_{2c}$ and $ph_{2t}$, she has to measure the phase difference between $ph_{2c}$-$ph_{1c}$, $ph_{2c}$-$ph_{1t}$, $ph_{2t}$-$ph_{1c}$, and $ph_{2t}$-$ph_{1t}$. Thus, even if Eve can measure the phase difference without destroying the individuals phase information, she has four phase difference values and she has to guess which of them represents the correct information.

## 4. Quantum protocols

The first application of the scheme shown in Fig. 1 is quantum key distribution. The DPSK-QKD protocol [10] can be directly implemented if Bob and Alice play the opposite roles as happens in traditional DPSK-QKD. Thus, Bob modulates randomly each pulse that reaches his place by applying the phases 0 or $\pi$. Alice, by its turn, is the one who has the interferometer placed at the coherent state output. The protocol rules are the same and its security is increased by the use of thermal states as explained before.

The second application is quantum secure direct communication where Bob sends to Alice in a secure way the bit string he wants. This is just a slight modification of the DPSK-QKD protocol. Since the thermal states protect the coherent states, Bob and Alice can divide the pulse sequence (emitted by Alice) in time slots having two pulses and code a bit in the phase difference of two consecutive coherent pulses. For example, if the phase difference is 0, a bit 0 is obtained and if the phase difference is $\pi$, the bit 1 is obtained. Furthermore, the time interval between two time slots is larger than the time interval between the pulses inside a time slot ($\tau$), hence, differently of the DPSK-QKD, there is no useful information between the second coherente pulse of the $n$-th time slot and the first coherent pulse of the ($n$+1)-th coherent pulse. Every time Alice gets detection, she will obtain, very likely, the bit encoded by Bob. Since weak coherent states are being used, in several times Alice will not have detection. She must inform these situations to Bob. Following, Alice sends a new sequence of pulses with a new polarisation codification and Bob, by its turn, retransmits the bits not obtained in the first round of the protocol. The process is repeated until Alice gets the complete information.

The third and last application is quantum secret sharing. Here, once more, a slot time of two consecutive pulses defines the bit value and, hence, there is no information between pulses of different time slots. The same setup shown in Fig. 1 is used but now there are several Bobs, each one having its own phase modulator and a secret bit sequence that defines the phase-shift values that each Bob must apply in the pulses sent by Alice. The individual Bob's sequences are built in such way that the difference of phase between the two pulses in a time slot is always 0 or $\pi$. At the end of the protocol, Alice is going to obtain a bit sequence that will allow her to perform a useful task. On the other hand, if any Bob does not use the correct phase-shifts, the difference of phases between pulses in a time slot may be different of 0 and $\pi$ or, maybe 0 when it should be $\pi$ or vice-versa. In this case, very likely, Alice will not get the correct final bit sequence. Hence, in order to Alice get the correct bit string, all Bobs must collaborate using correctly their phase modulators according to their secrets.

## 5. Experimental results

In order to show to feasibility of the optical setup proposed, we realized a proof-of-principle experiment, implementing the optical setup shown in Fig. 1 without Bob's phase modulator and using CW light. Our goal is to check the security of the proposed scheme. The optical channel used was a 1 km single-mode optical fibre. We used a home-made single-photon detector based on the avalanche photodiode

PGA-400, from Princenton Lightwave Inc. The output signal of the single-photon detector was directly plugged in a spectrum analyzer and the electrical power was measured. Two laser diodes operating at 1550 nm were used. The thermal light was produced operating one of the lasers well below the threshold while the coherent state was produced operating the other laser well above the threshold. The mean photon number used for both quantum states was $\mu$~0.1. The electrical power in a fixed band (central frequency = 70 MHz and RBW = 2 MHz) was measured in four situations: I) $\theta_1=\theta_2=0$. II) $\theta_1=\theta_2=\pi/2$. III) $\theta_1=0$ and $\theta_2=\pi/4$. IV) Complete absence of light. The cases I and II show the expected measured power value when there are not attacks. The case III simulates an intercept-resend attack in which Eve changes the coherent and thermal modes with probability 50%. At last, in case IV the power measured is produced by avalanches caused by dark counts. These results can be seen in Fig. 2.

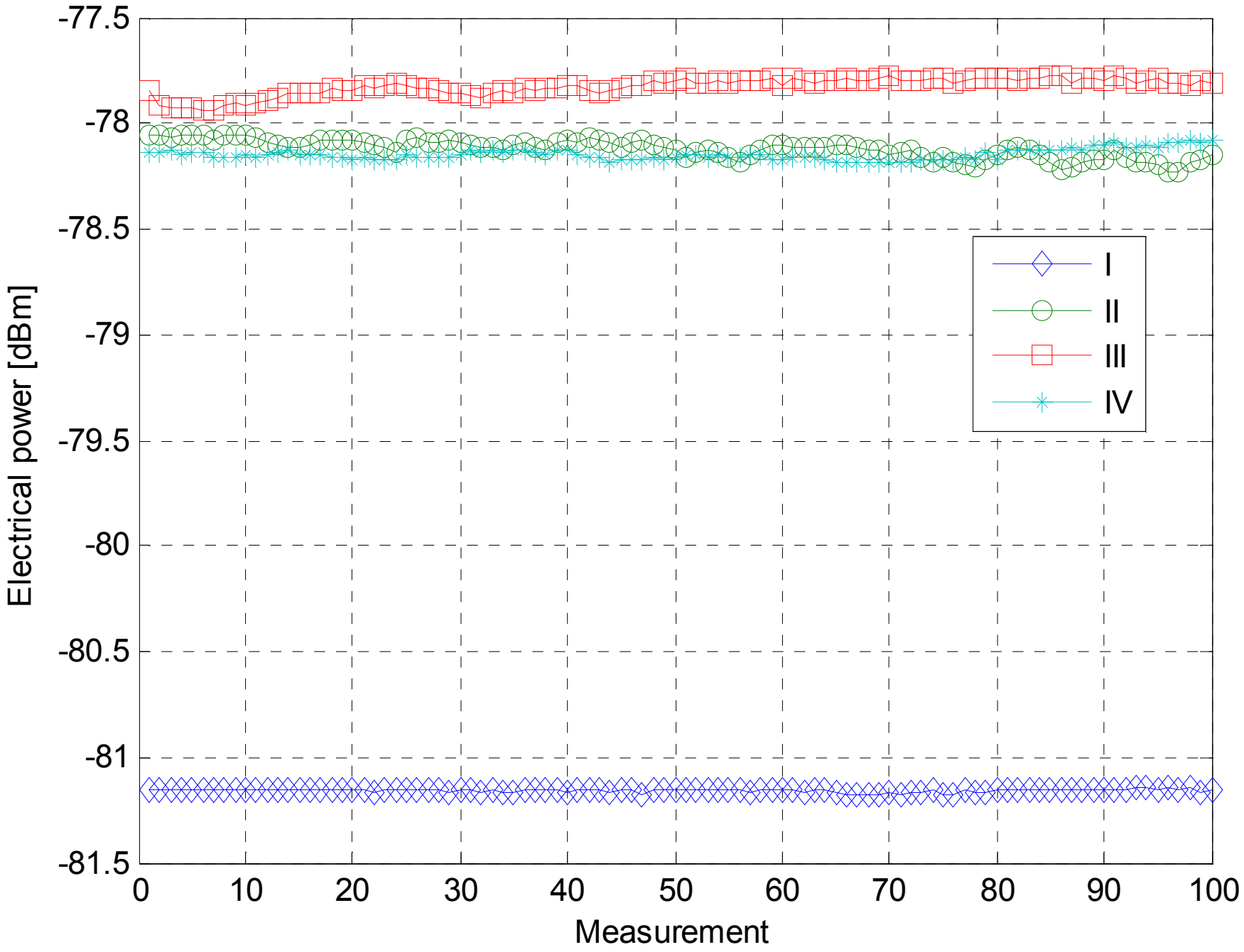

Figure 2 - Electrical power measured by a spectrum analyzer. A hundred values were measured in order to show the stability of the measurement along the time. I) $\theta_1=\theta_2=0$. II) $\theta_1=\theta_2=\pi/2$. III) $\theta_1=0$ and $\theta_2=\pi/4$. IV) Absence of light.

As can be observed in Fig. 2, when there are coherent and thermal states at the thermal output (case III), what happens in the intercept-resend attack, Alice clearly measures a power value different from the expected value (cases I and II).

## 6. Conclusions

Although we have discussed only QKD, QSDC and QSS protocols, our setup can be used in any quantum protocol where the secure transmission of coherent states is required. At last, since our optical scheme is secure, easily implementable and it can support different quantum protocols, we believe that it is a step-forward in the implementation of multi-service quantum networks.

## Acknowledgements

This work was supported by the Brazilian agencies CAPES and CNPq via Grant no. 303514/2008-6. The experimental work was realized at LATIQ and NUCEMA/NUTEC(SECITECE) laboratories. Also, this work was performed as part of the Brazilian National Institute of Science and Technology for Quantum Information.